\documentclass[12pt,english,aps,preprint]{article}
\usepackage[T1]{fontenc}
\usepackage[latin9]{inputenc}
\usepackage[a4paper]{geometry}
\usepackage{array}
\usepackage{amstext}
\usepackage{graphicx}
\usepackage{setspace}
\usepackage{esint}
\makeatletter

\newcommand{\noun}[1]{\textsc{#1}}
\providecommand{\tabularnewline}{\\}
\newcommand{\lyxaddress}[1]{
\par {\raggedright #1
\vspace{1.4em}
\noindent\par}
}

\makeatother

\usepackage{babel}

\begin{document}

\title{
\begin{flushright}
 \small MZ-TH/10-33
\end{flushright}
\vspace{2cm}
Exact Curie temperature for the Ising model on Archimedean and Laves lattices}

\author{Alessandro Codello}

\maketitle

\lyxaddress{\begin{center}
Institut f\"ur Physik, Johannes Gutenberg-Universit\"at, Mainz
\par\end{center}}
\begin{abstract}
Using the Feynman-Vdovichenko combinatorial approach to the two dimensional
Ising model, we determine the exact Curie temperature for all two dimensional
Archimedean lattices. By means of duality, we extend our results to
cover all two dimensional Laves lattices. For those lattices where
the exact critical temperatures are not exactly known yet, we compare
them with Monte Carlo simulations.
\end{abstract}

\section{Introduction}

As a solvable example of a statistical system exhibiting a second
order phase transition, the Ising model has been of central importance
in the development of statistical mechanics and still continues to
play a key role as test ground for new theoretical and computational
methods.

The existence of a Curie point, at which a transition from the paramagnetic
phase to the ferromagnetic phase takes place, as the temperature is
lowered, was first proven to exist by Peierls \cite{Peierls_1936}
in the case of the square lattice Ising model. Successively Kramers
and Wannier \cite{Kramers_Wannier_1941} calculated the value of the
critical temperature using duality and the property of the square
lattice of being self-dual. With the first calculation of the zero
field free energy by Onsager
\cite{Onsager_1944,Kaufman_1949} and the calculation of the spontaneous
magnetization by Yang \cite{Yang_1952}, the second order character
of the phase transition was established. This successful attempt to
solve the Ising model on a square lattice was carried over using transfer matrix techniques.
Along the same path, Wannier was first to study the triangular and
hexagonal (also called honeycomb) lattice Ising model, in particular
in relation to antiferromagnetism studies \cite{Wannier_1950}. The
first non regular lattice zero field free energy was calculated for
the ``Kagome'' lattice by Kano and Naya \cite{Kano_Naya_1953}.
Successively the ``extended Kagome'' lattice was studied by Soyzi
\cite{Soyzi_1972}. The ``Union Jack'' lattice and the dual, the {}``Bathroom-tile''
lattice, were covered by \cite{Utiyama_1951,Vaks_Larkin_Ovchinnikov_1966}.
Other lattices were investigated in \cite{Urumov_2002,Strecka_2005}.

A combinatorial approach to solve the Ising model was proposed by
Kac and Ward \cite{Kac_Ward_1952}. It is based on the possibility
to map the high temperature expansion of the partition function onto
a particular random-walk problem on the lattice. Feynman \cite{Feynman_1972}
and Vodvicenko \cite{Vodvicenko_1965} showed how to recover the Onsager
solution by this method. The above mapping is based on a topological
theorem about loops on two dimensional lattices proven by Sherman
\cite{Sherman_1960}.

Despite of many efforts, the free energy in presence of a non-zero
external magnetic field has not yet been calculated exactly. Some
results are available only at criticality $T=T_{c}$ or for weak magnetic
field $B\rightarrow0$ \cite{Fonseca_Zamolodchikov_2003}. Still nowadays
the Ising model is an active area of research.

In this prospect any exact analytical result can be useful. In this
paper we use the combinatorial method to calculate the Curie temperatures
for the Ising model on all Archimedean lattices and all their duals,
the Laves lattices. This method was first used by Thompson and Wardrop
\cite{Thompson_Wardrop_1974} to calculate the critical temperatures
for two Archimedean lattices, here we calculate the Curie temperature
for all of them, including three values which were known by now only
approximately from Monte Carlo simulations.

\section{Archimedean and Laves lattices}

Archimedean lattices are uniform tilings of the plane in which all
the faces are regular polygons and the symmetry group acts transitively
on the vertices. It follows that all vertices are equivalent and have
the same coordination number $z$. Also, all vertices are shared by
the same set of polygons and thus we can associate to each Archimedean
lattice a set of integers $(p_{1},p_{2},...)$ indicating, in cyclic
order, the polygons meeting at a given vertex \cite{Grunbaum_Shephard_1989}.
When a polygon appears more than one time consecutively, for example
as in $(...,p,p,...)$, we abbreviate the notation writing $(...,p^{2},...)$.%
\begin{figure*}[t]
\begin{centering}
\includegraphics[scale=0.38]{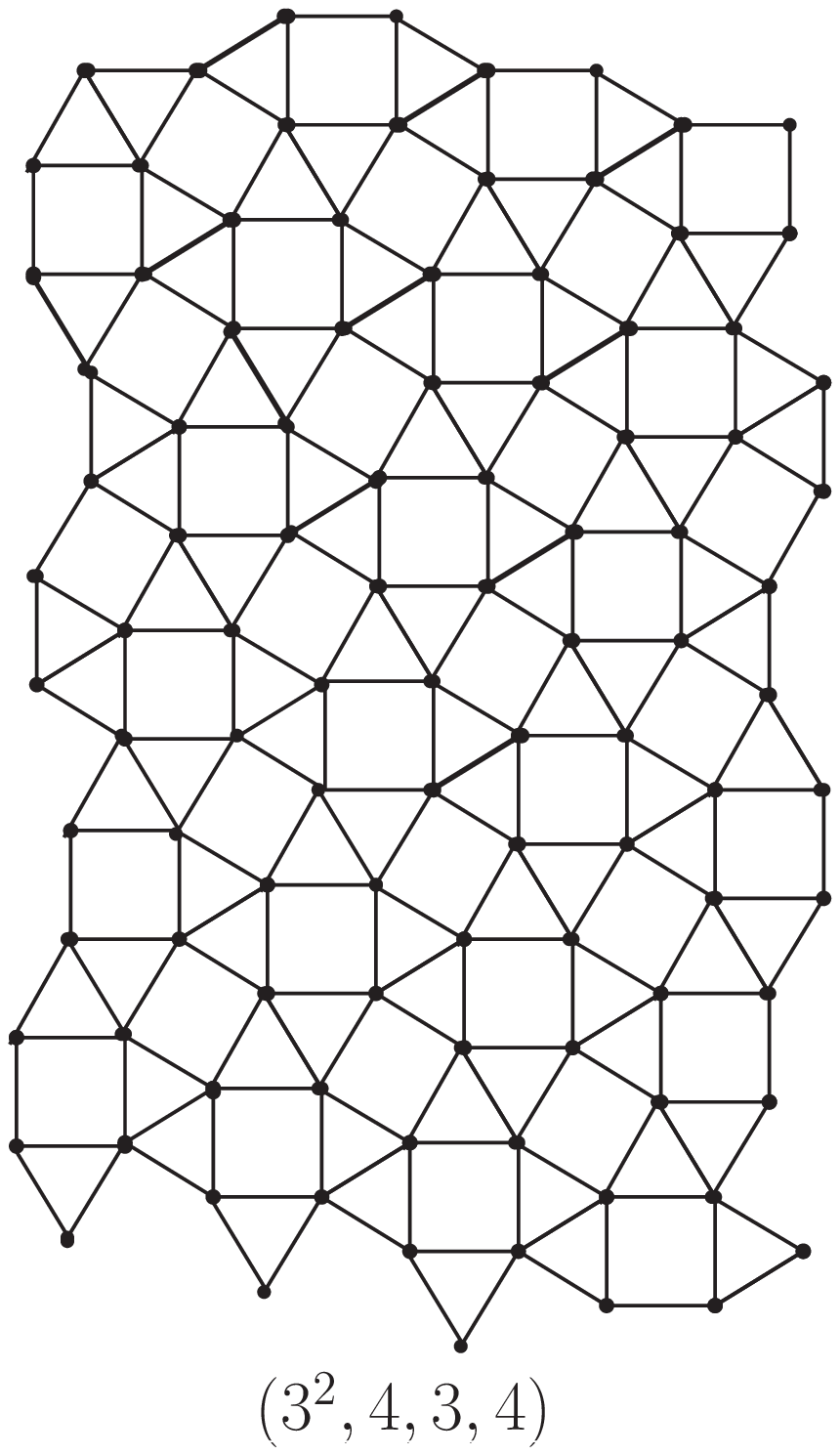}
\includegraphics[scale=0.38]{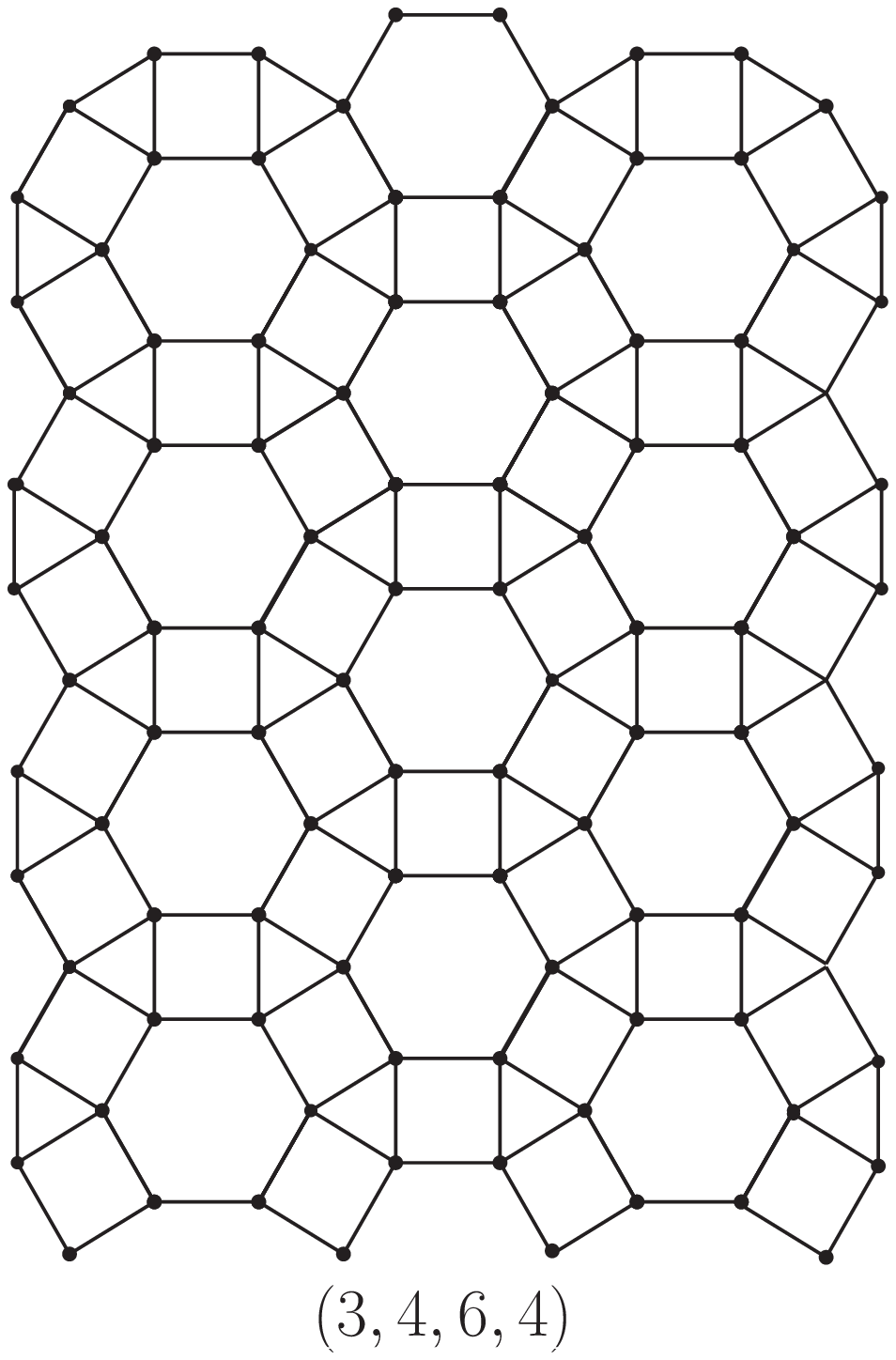}
\includegraphics[scale=0.4]{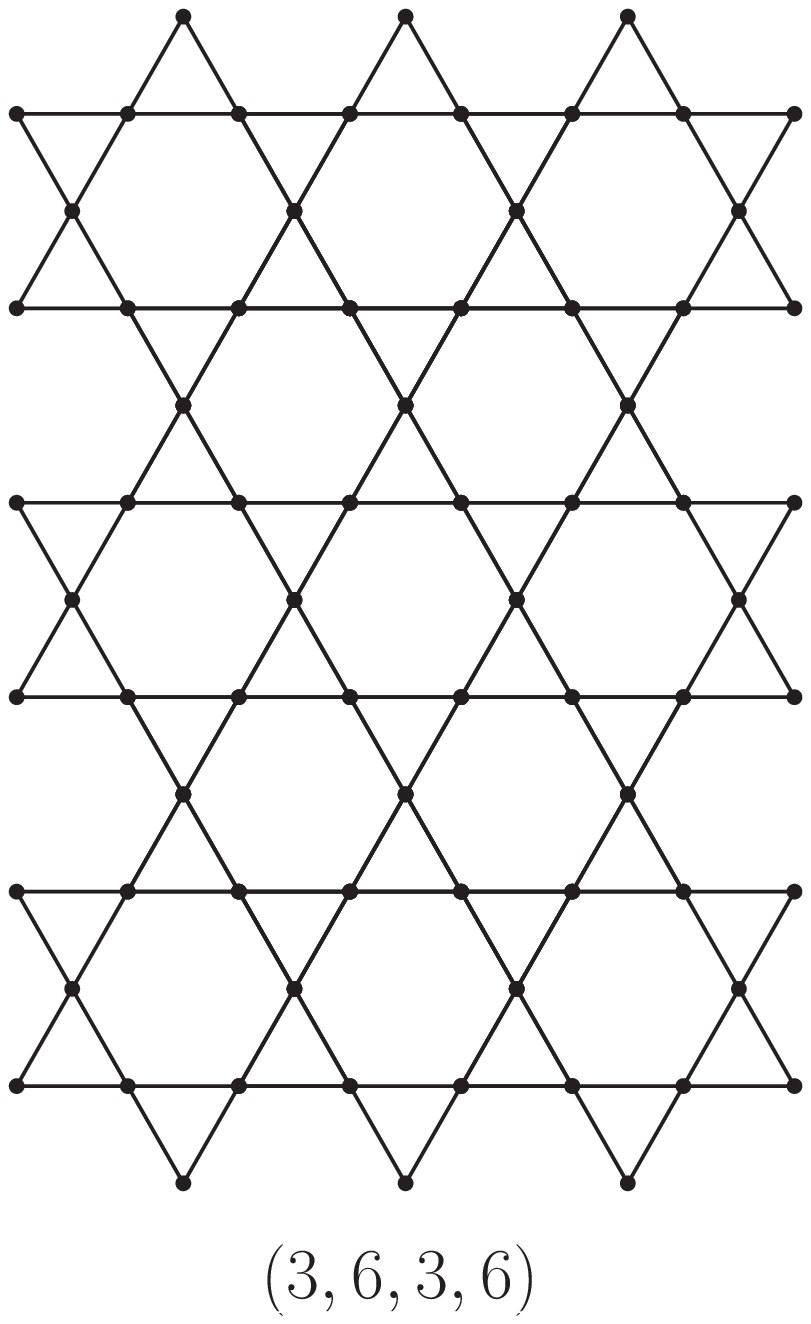}
\includegraphics[scale=0.4]{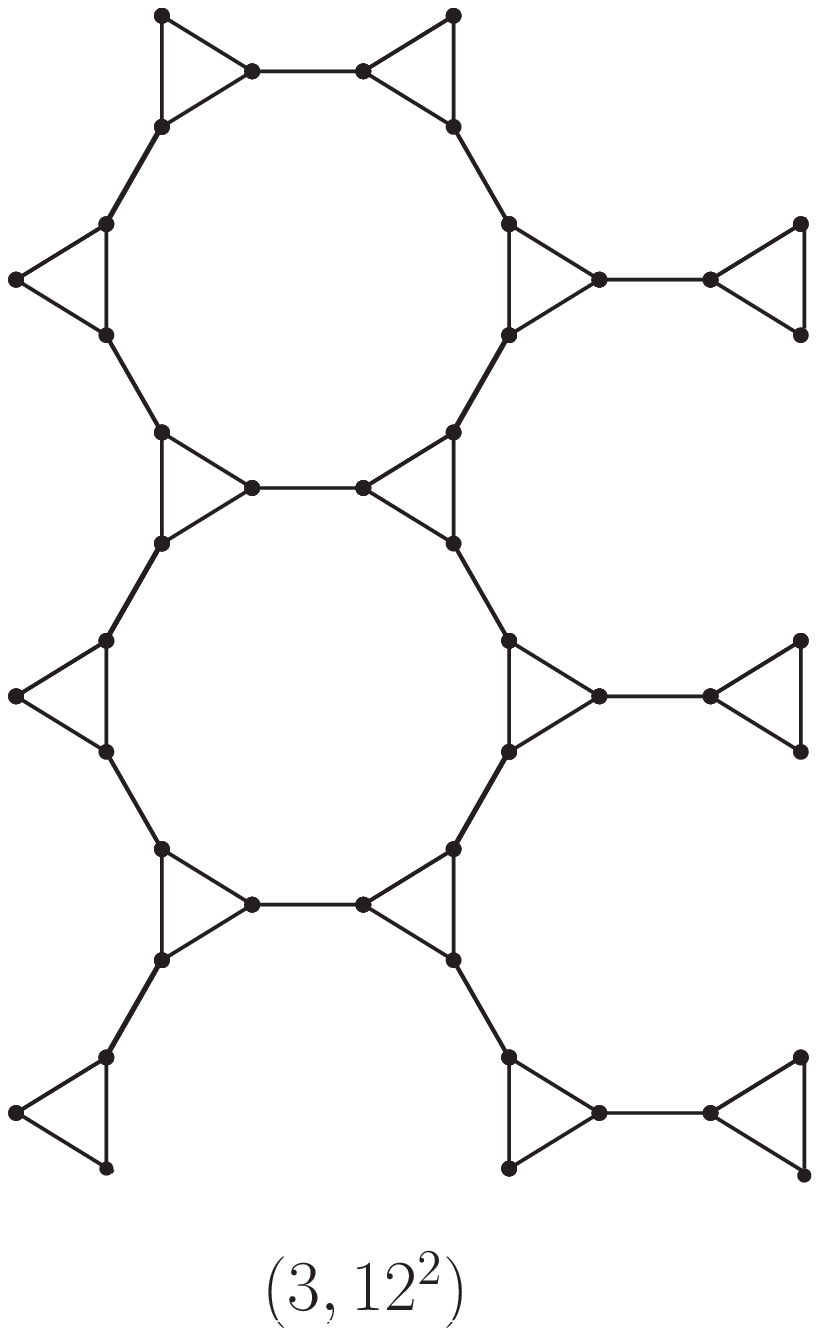}\\
\includegraphics[scale=0.45]{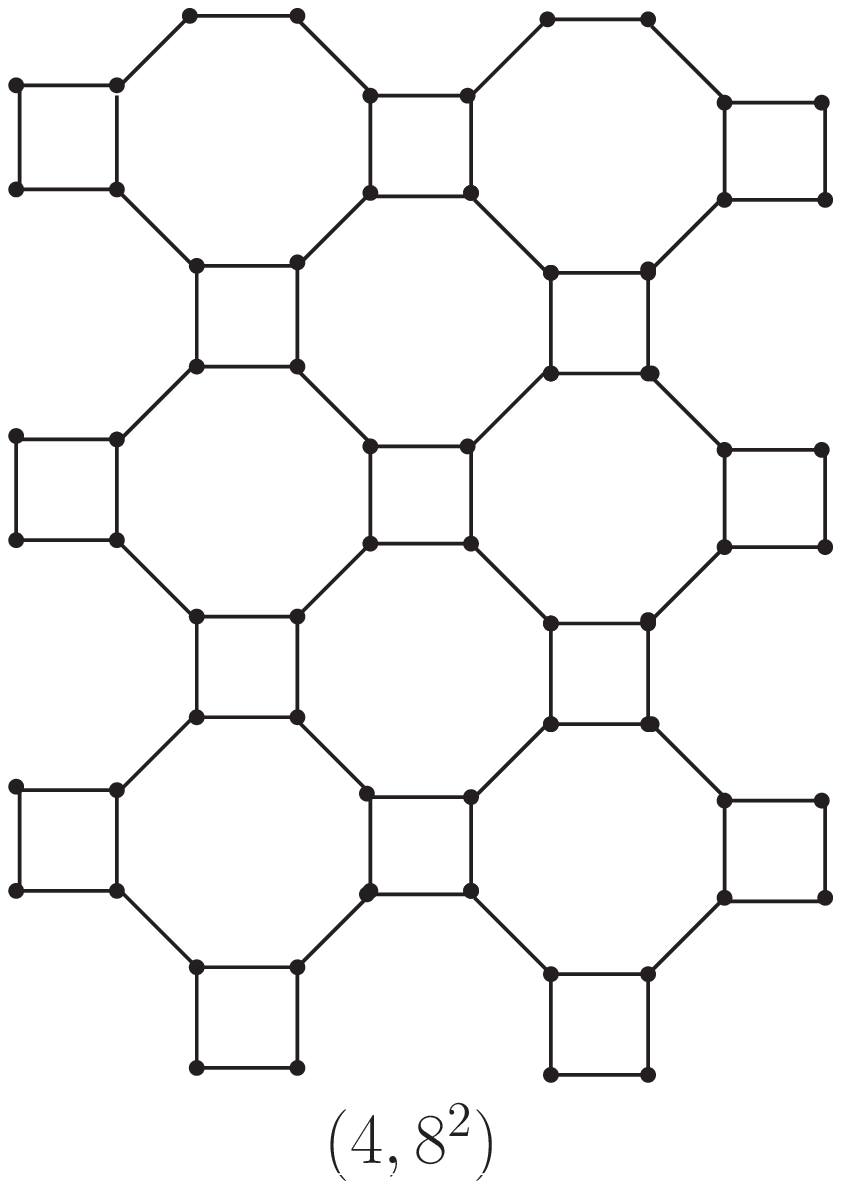}
\includegraphics[scale=0.42]{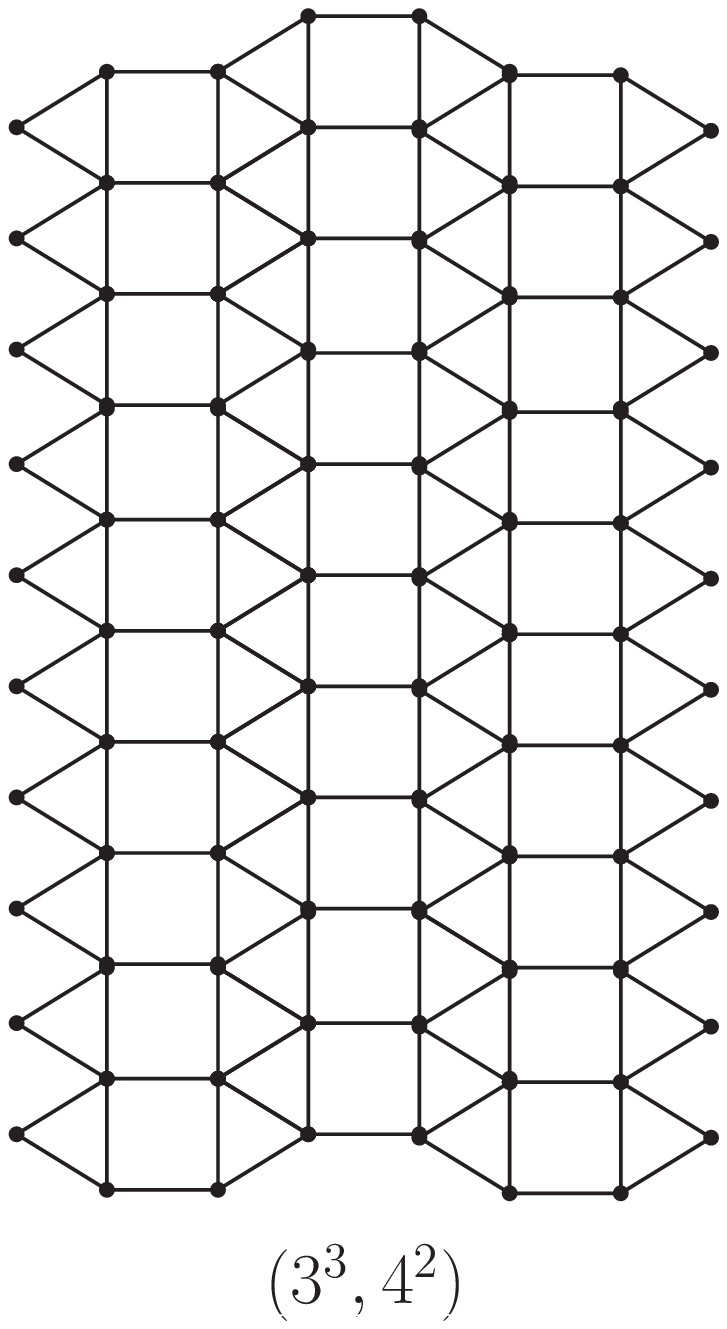}
\includegraphics[scale=0.38]{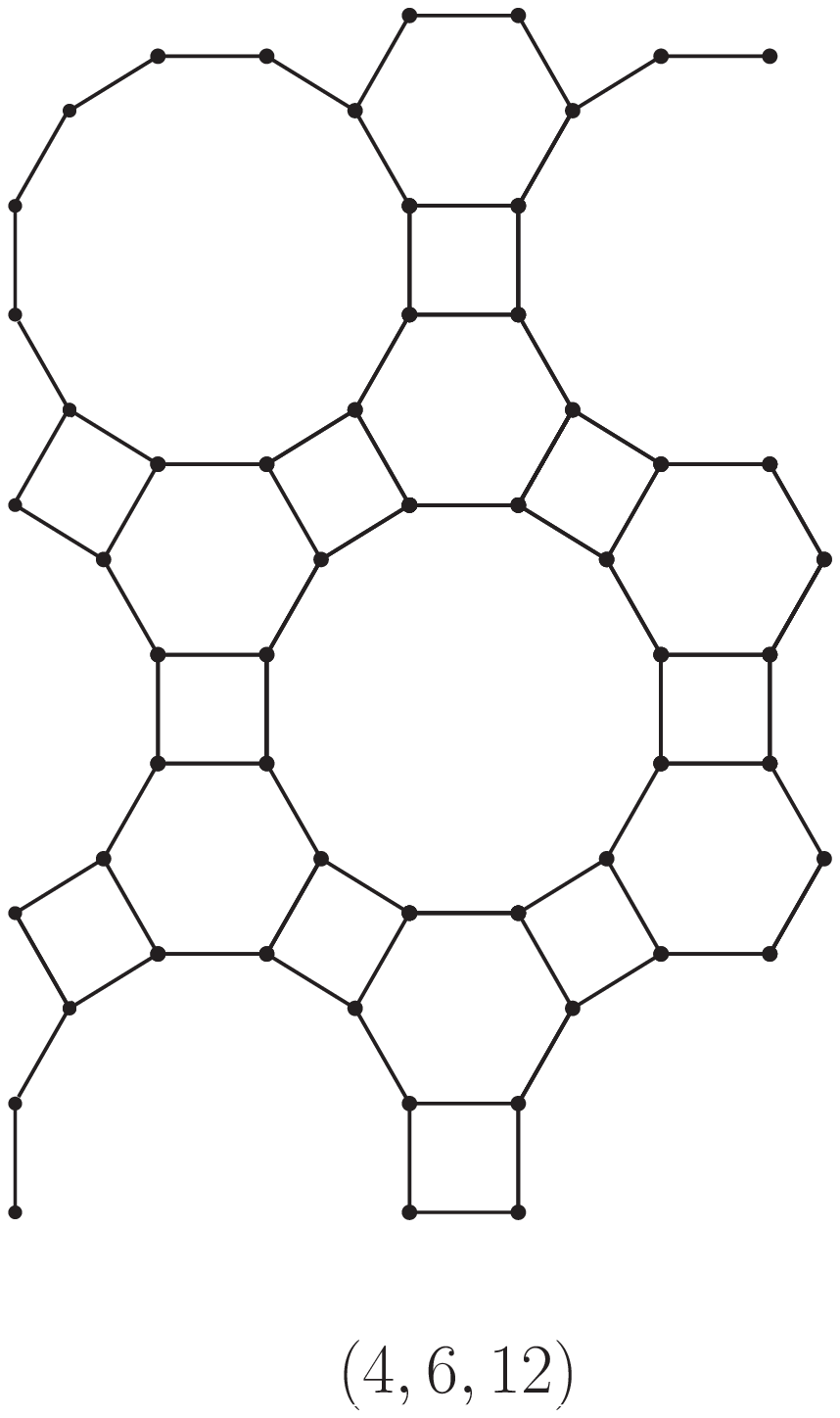}
\includegraphics[scale=0.52]{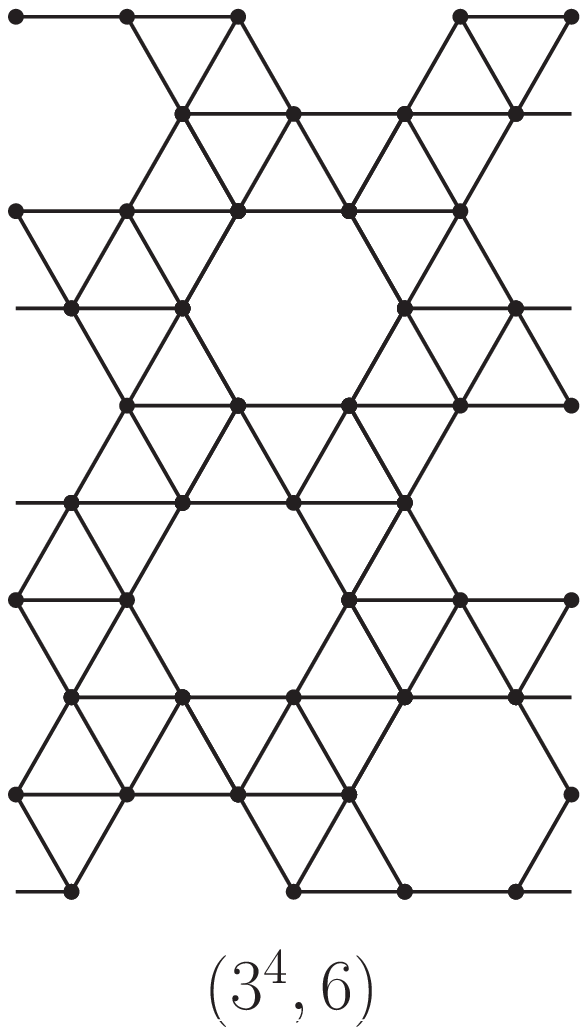}
\par\end{centering}

\caption{The eight uniform tilings of the plane made with more than one regular polygon.}

\end{figure*}
 In this way, the uniform tilings of the plane, made with squares,
triangles and hexagons, are indicated respectively by $(4^{4})$, $(3^{6})$
and $(6^{3})$.

It is possible to show, that apart the regular tilings just listed,
there are eight other uniform tilings of the plane. This are all constructed 
using more than one regular polygon and are shown in Figure 1. In total we have
eleven different two dimensional Archimedean lattices. Some of this lattices
 have names: ``Kagome''
$(3,6,3,6)$, ``extended Kagome'' $(3,12^{2})$ and ``Bathroom-tile''
$(4,8^{2})$, while the others have no particular name. Note that
the lattice $(3^{4},6)$ is the only non-specular one.

The dual lattice $\Lambda^{*}$ of a given lattice $\Lambda$ is constructed
associating a vertex of $\Lambda^{*}$ to each face of $\Lambda$.
The duals of the Archimedean lattices are called Laves lattices. They
are tilings of the plane made with one non-regular polygon and are
indicated by the notation $[p_{1},p_{2},...]$. The square lattice
is self-dual $[4^{4}]=(4^{4})$ (this property was used by Kramers
and Wannier to first calculate the critical temperature for this lattice),
while the triangular and hexagonal lattices are dual of each other
$[3^{6}]=(6^{3})$, $[6^{3}]=(3^{6})$. Of the duals of the other uniform
lattices, three are triangular tilings, two are rhomboidal tilings
and three are pentagonal tilings. The dual of the Kagome lattice $[3,6,3,6]$
is called ``Dice lattice'' and the lattice $[4,8^{2}]$, dual to the ``Bathroom tile`` lattice, is called
``Union Jack'' lattice.

\section{Ising model}

We briefly review the definition of the model and the basic steps involved
in the combinatorial solution of it. At every lattice site $i$ a spin variable
$\sigma_{i}\in\{-1,1\}$ is located and a microstate is defined by
a spin configuration $\{\sigma\}$. Nearest neighbor interactions
are assumed so that the energy of a given spin configuration is given
by
\begin{equation}
E\{\sigma\}=-J\sum_{\left\langle i,j\right\rangle }\sigma_{i}\sigma_{j}-B\sum_{i}\sigma_{i}\,.\label{defising}
\end{equation}
If $J>0$ the interaction is ferromagnetic, while it is antiferromagnetic
if $J<0$. For the aim of calculating the Curie temperature, we can
set to zero the external applied magnetic field $B=0$. The partition
function is the sum over all spin configurations, weighted by the Boltzman-Gibbs
factor
\begin{equation}
Z_{\Lambda}=\sum_{\{\sigma\}}\exp \left\{K\sum_{\left\langle i,j\right\rangle }\sigma_{i}\sigma_{j}\right\}\,.\label{part}
\end{equation}
We defined $K=J/k_{B}T$ where $k_{B}$ is the Boltzmann constant.

Using the high temperature expansion, the partition function (\ref{part}),
can be rewritten as:
\begin{equation}
Z_{\Lambda}=2^{N}\left(\cosh K\right)^{N_{E}}\Phi_{\Lambda}(v)\,.\label{2.1}
\end{equation}
In (\ref{2.1}) we defined $v=\tanh K$, while $N$ is the number of lattice sites and $N_{E}=\frac{z}{2}N$
is the number of links per lattice site. The function $\Phi_{\Lambda}(v)$
is the generating function of the numbers $G_{n}$ (we define $G_{0}=1$), which count the
graphs of length $n$ with even vertices  that
can be drawn on the lattice $\Lambda$. In this way, the problem of
solving the Ising model on $\Lambda$ is reduced to the
combinatorial problem of counting even closed graphs on $\Lambda$.

If we don't care about the fact that we have to count only over closed
graphs with even vertices, we can use the standard combinatorial property
that the sum over all (possibly disconnected) closed graphs, is given
by the exponential of the sum over only connected closed graphs. But
connected closed graphs can be seen as closed random-walk paths. If
we find a way to consider in the sum only closed paths with even vertices,
we have a way to calculate the generating function $\Phi_{\Lambda}(v)$
simply by counting closed random-walks paths. 

Feynman \cite{Feynman_1972} and Vodvicenko \cite{Vodvicenko_1965}
introduced a trick to count only over paths with even vertices: weight
the random-walk with a complex amplitude $\alpha=e^{i\theta/2}$ at
every turn of angle $\theta$ the random-walker makes along his trajectory.
In this way all contributions coming from non-even paths delete each
other in the sum, and the mapping from the high temperature expansion
to the random-walk problem works out correctly (see \cite{Kardar_2007,Mussardo_2009}
for more details). This results have a rigorous base on a topological
theorem proven by Sherman \cite{Sherman_1960} and were further investigated
by Morita \cite{Morita_1985}.

After introducing the appropriate amplitudes, which depend on the
properties of the lattice $\Lambda$, we need to count all connected
closed graphs of length $n$ that can be draw on $\Lambda$. This
number is equal to the number of random-walks of length $n$ starting
from all the lattice sites
\begin{equation}
\frac{1}{2n}\sum_{\mathbf{x}_{0},i_{0}}N_{i_{0}}(\mathbf{x}_{0},\mathbf{x}_{0},n)\,,\label{gconn}
\end{equation}
where $N_{i}(\mathbf{x},\mathbf{x}_{0},n)$ is the number of walks
arriving at site $\mathbf{x}$ from direction $i$ at the $n$-th
step starting from site $\mathbf{x}_{0}$. The sum is over all possible
starting points and the factor $\frac{1}{2n}$ exclude the over-counting
due to the fact that there are two directions in which a graph can
be travelled and the fact that each graph can be seen as a random-walk
starting from all of the sites of the lattice reached by the random-walker.

The random-walk problem can be written in terms of the master equation%
\begin{table}[t]
\begin{singlespace}
\begin{centering}
\begin{tabular}{llclclc}
\hline 
$\Lambda$ &  & $z$ &  & $p$ &  & $m$\tabularnewline
\hline 
$(3,12^{2})$ &  & $3$ &  & $6$ &  & $18$\tabularnewline
$(4,6,12)$ &  & $3$ &  & $12$ &  & $36$\tabularnewline
$(4,8^{2})$ &  & $3$ &  & $4$ &  & $12$\tabularnewline
$(6^{3})$ &  & $3$ &  & $2$ &  & $6$\tabularnewline
\hline 
$(3,4,6,4)$ &  & $4$ &  & $6$ &  & $24$\tabularnewline
$(4^{4})$ &  & $4$ &  & $1$ &  & $4$\tabularnewline
$(3,6,3,6)$ &  & $4$ &  & $3$ &  & $12$\tabularnewline
\hline 
$(3^{4},6)$ &  & $5$ &  & $6$ &  & $30$\tabularnewline
$(3^{3},4^{2})$ &  & $5$ &  & $2$ &  & $10$\tabularnewline
$(3^{2},4,3,4)$ &  & $5$ &  & $4$ &  & $20$\tabularnewline
\hline 
$(3^{6})$ &  & $6$ &  & $1$ &  & $6$\tabularnewline
\hline
\end{tabular}
\par\end{centering}
\end{singlespace}

\caption{Coordination number $z$, number of lattice sites per unit cell $p$
and size $m$ of the random-walk matrix $\mathbf{W}_{\Lambda}$ for
each Archimedean lattice.}

\end{table}
\begin{equation}
N_{i}(\mathbf{x},\mathbf{x}_{0},n)=\sum_{\mathbf{x}',j}W_{i,j}(\mathbf{x},\mathbf{x}')N_{j}(\mathbf{x}',\mathbf{x}_{0},n-1)\,,\label{master}\end{equation}
with boundary conditions $N_{i}(\mathbf{x},\mathbf{x}_{0},0)=\delta_{\mathbf{x},\mathbf{x}_{0}}\delta_{i,i_{0}}$.
Equation (\ref{master}) is solved by $N_{i}(\mathbf{x},\mathbf{x}_{0},n)=W_{i,i_{0}}^{n}(\mathbf{x},\mathbf{x}_{0})$.
From equation (\ref{gconn}) we find, in the limit $N\rightarrow\infty$, the explicit form for the generating function:\[
\Phi_{\Lambda}(v)=\exp\left\{ \frac{N}{2}\int\frac{d^{2}k}{(2\pi)^{2}}\log\,\textrm{det}\left[\mathbf{1}-v\mathbf{W}_{\Lambda}(k_{x},k_{y})\right]\right\} \,,\]
where the integration is over the region $0\leq k_{x}\leq2\pi$ and
$0\leq k_{y}\leq2\pi$. The free energy for spin $f_{\Lambda}=F_{\Lambda}/N$
for the lattice $\Lambda$, can finally be written as:\begin{equation}
f_{\Lambda}/k_{B}T=-p\log2-\frac{z}{2}\log(1+v^{2})-\frac{1}{2}\int\frac{d^{2}k}{(2\pi)^{2}}\textrm{log}\,\textrm{det}\left[\mathbf{1}-v\mathbf{W}_{\Lambda}(k_{x},k_{y})\right]\,.\label{logdet}\end{equation}
The matrices $\mathbf{W}_{\Lambda}(k_{x},k_{y})$ are $m\times m$
matrices with $m=zp$. Here $z$ is the coordination number of the
lattice $\Lambda$, while $p$ is the number of lattice sites in the
unit cell of $\Lambda$. The values of $m$, $z$ and $p$
for the Archimedean lattices $\Lambda$ are reported in Table 1.

\section{Critical temperatures}

At the phase transition the free energy becomes singular, from equation
(\ref{logdet}) we see that this can happen only if the argument of
the logarithm becomes zero. For each $\Lambda$ this can happen only
for $k_{x}=k_{y}=0$. To every lattice we associate the polynomials:%
\begin{table*}
\begin{onehalfspace}
\begin{centering}
\begin{tabular}{lllll}
\hline 
$z$ &  & $\Lambda$ &  & $P_{\Lambda}(v)$\tabularnewline
\hline
$3$ &  & $(3,12^{2})$ &  & $(1+v)^{4}(1-2v+3v^{2}-2v^{3}-2v^{4})^{2}$\tabularnewline
 &  & $(4,6,12)$ &  & $(1+2v^{2}+5v^{4})^{2}(1-2v^{2}+2v^{4}-10v^{6}+v^{8})^{2}$\tabularnewline
 &  & $(4,8^{2})$ &  & $(1-v-v^{2})(1+v+2v^{2})(1-4v^{3}-v^{4})$\tabularnewline
 &  & $(6^{3})$ &  & $(1-3v^{2})^{2}$\tabularnewline
\hline
$4$ &  & $(3,4,6,4)$ &  & $(1-v^{2})^{2}(1+v^{2})^{6}(1-4v^{2}-6v^{4}-4v^{6}+v^{8})$\tabularnewline
 &  & $(3,6,3,6)$ &  & $(1-v^{2})^{2}(1-4v^{2}-6v^{4}-4v^{6}+v^{8})$\tabularnewline
 &  & $(4^{4})$ &  & $(1-2v-v^{2})^{2}$\tabularnewline
\hline 
$5$ &  & $(3^{4},6)$ &  & $(1+v)^{8}(1+3v^{2})^{2}(1-4v+7v^{2}-12v^{3}+3v^{4}-3v^{6})^{2}$\tabularnewline
 &  & $(3^{3},4^{2})$ &  & $(1+v)^{2}(1-3v)^{2}(1+v^{2})^{2}$\tabularnewline
 &  & $(3^{2},4,3,4)$ &  & $(1+v)^{4}(1-2v-v^{2}-4v^{3}-9v^{4}+6v^{5}-7v^{6})^{2}$\tabularnewline
\hline 
$6$ &  & $(3^{6})$ &  & $(1+v)^{2}(1-4v+v^{2})^{2}$\tabularnewline
\hline
\end{tabular}
\par\end{centering}
\end{onehalfspace}

\caption{The polynomials $P_{\Lambda}(v)$ as defined in equation (\ref{poly}), for
each Archimedean lattice.}

\end{table*}
\begin{equation}
P_{\Lambda}(v)=\textrm{det}\left(\mathbf{1}-v\mathbf{W}_{\Lambda}(0,0)\right)\,.\label{poly}\end{equation}
The critical values of the high temperature parameter $0<v_{c}<1$
are the real solution of the equation \begin{equation}
P_{\Lambda}(v)=0\,.\label{Curie}\end{equation}
The polynomials $P_{\Lambda}(v)$ for the various Archimedean lattices
are listed in Table 2. For all the lattices we considered, we found
only one real solution of equation (\ref{Curie}) in the interval
$(0,1)$. In all, but in the cases of the lattices $(3^{4},6)$ and
$(3^{2},4,3,4)$, equation (\ref{Curie}) has an explicit algebraic
solution. The critical values $v_{c}$ are listed in the collum 3 of Table 3.%
\begin{table*}[t]
\begin{spacing}{1.7}
\begin{centering}
\begin{tabular}{llllllll}
\hline 
$z$ &  & $\Lambda$ &  & $v_{c}$ & $k_{B}T_{c}/J$ &  & $k_{B}T_{c}^{*}/J$ \tabularnewline
\hline 
$3$ &  & $(3,12^{2})$ &  & $-\frac{1}{4}-\frac{\sqrt{3}}{4}+\frac{1}{2}\sqrt{3+\frac{5\sqrt{3}}{2}}$ & $1.2315$ &  & $5.0071$ \tabularnewline
 &  & $(4,6,12)$ &  & $\sqrt{\frac{5+3\sqrt{3}-\sqrt{44+26\sqrt{3}}}{2}}$ & $1.3898$ &  & $4.1363$ \tabularnewline
 &  & $(4,8^{2})$ &  & $-1-\frac{1}{\sqrt{2}}+\sqrt{\frac{5+4\sqrt{2}}{2}}$ & $1.4387$ &  & $3.9310$ \tabularnewline
 &  & $(6^{3})$ &  & $\frac{1}{\sqrt{3}}$ & $1.5186$ &  & $3.6410$ \tabularnewline
\hline 
$4$ &  & $(3,4,6,4)$ &  & $\frac{1}{2}-\sqrt{\frac{\sqrt{3}}{2}}+\frac{\sqrt{3}}{2}$ & $2.1433$ &  & $2.4055$ \tabularnewline
 &  & $(3,6,3,6)$ &  & $\frac{1}{2}-\sqrt{\frac{\sqrt{3}}{2}}+\frac{\sqrt{3}}{2}$ & $2.1433$ &  & $2.4055$ \tabularnewline
 &  & $(4^{4})$ &  & $\sqrt{2}-1$ & $2.2692$ &  & $2.2692$ \tabularnewline
\hline 
$5$ &  & $(3^{4},6)$ &  & $0.344296$ & $2.7858$ &  & $1.8757$ \tabularnewline
 &  & $(3^{3},4^{2})$ &  & $\frac{1}{3}$ & $2.8854$ &  & $1.8205$ \tabularnewline
 &  & $(3^{2},4,3,4)$ &  & $0.32902$ & $2.9263$ &  & $1.7992$ \tabularnewline
\hline 
$6$ &  & $(3^{6})$ &  & $2-\sqrt{3}$ & $3.6410$ &  & $1.5186$ \tabularnewline
\hline
\end{tabular}
\par\end{centering}
\end{spacing}

\caption{For each Archimedean lattice, the exact critical value of the high temperature parameter $v_{c}$,
of the Curie temperature $T_{c}$ and of the dual Curie temperature $T_{c}^{*}$, are given.
The Curie temperatures $T_{c}$ and $T_{c}^{*}$ are calculated from $v_{c}$
using equations (\ref{8}) and (\ref{9}). 
The values for the lattices $(4,6,12)$, $(3,4,6,4)$ and $(3^{4},6)$ were not yet known
exactly and agree, within errors, with the numerical values of \cite{Lima_Mostowicz_Malarz_2010,Malarz_Zborek_Wrobel_2005}.
The critical values for the lattices $(3^{3},4^{2})$ and $(3^{2},4,3,4)$ agree with the
exact results of \cite{Thompson_Wardrop_1974}. 
Note that the lattices $(3,4,6,4)$ and $(3,6,3,6)$ have the same
critical temperature.}

\end{table*}

The critical temperatures $T_{c}$ are related to the high temperature
parameter $v_{c}$ by the following relation:
\begin{equation}
T_{c}=\frac{J/k_{B}}{\textrm{arctanh}v_{c}}\,.\label{8}
\end{equation}
The values we found are given in the collum 4 of Table 3. The exact Curie temperatures
for all the lattices, except for $(4,6,12)$, $(3,4,6,4)$
and $(3^{4},6)$, were already known in the literature. 
In particular, they can be found in the following references: $(4^{4})$ in \cite{Kramers_Wannier_1941,Onsager_1944,Kaufman_1949}, $(3^{6})$ and $(6^{3})$ in \cite{Wannier_1950}, $(3,12^{2})$ in \cite{Soyzi_1972,Matveev_Shrock_1995}, $(4,8^{2})$ in \cite{Utiyama_1951,Vaks_Larkin_Ovchinnikov_1966,Matveev_Shrock_1995}, $(3,6,3,6)$ in \cite{Kano_Naya_1953,Matveev_Shrock_1995}, $(3^{3},4^{2})$ in \cite{Thompson_Wardrop_1974} and $(3^{2},4,3,4)$ in \cite{Thompson_Wardrop_1974,Strecka_2005,Urumov_2002}.

We have compared our exact results for the lattices $(4,6,12)$, $(3,4,6,4)$
and $(3^{4},6)$ with the numerical values available
from Monte Carlo simulations \cite{Lima_Mostowicz_Malarz_2010,Malarz_Zborek_Wrobel_2005} and they agree within numerical errors. This is an independent confirmation of the correctness of our analytical calculations for this new critical values.

The critical temperatures for the lattices $(3^{2},4,3,4)$ and $(3^{3},4^{2})$
were calculated in \cite{Thompson_Wardrop_1974} using the same combinatorial
approach we are using, and agree with our findings. The fact that we used lattices
where all edges are of unit length, contrary to \cite{Thompson_Wardrop_1974}
who used a different ``embedding'' of the lattices by constructing
them from a square lattice adding diagonal links, shows that the critical
temperatures depend only on the topology of the lattice and not on
the particular representation of it. In fact, different embeddings
of the same lattice lead to different random-walk matrices $\mathbf{W}_{\Lambda}(0,0)$.
This give rise to different polynomials $P_{\Lambda}(v)$, that instead
have the same roots in the interval $(0,1)$. The fact that the Curie
temperatures depend only on the topology of the lattice is understood
from the fact that the function $\Phi_{\Lambda}(v)$, which counts the number
of closed even graphs, is a topological property of the lattice which
is independent of the embedding.

Note that the lattices $(3,4,6,4)$ and $(3,6,3,6)$ have the same
Curie temperature. This comes from the fact that the relative polynomials
$P_{(3,4,6,4)}(v)$ and $P_{(3,6,3,6)}(v)$ have a common factor with
root in the interval $(0,1)$. This may suggest that the critical
probabilities $p_{c}$ for correlated percolation (also know as bootstrap
percolation) on this lattices are equal.

Using the critical values for the high temperature parameter $v_{c}$
found for the Archimedean lattices, we can calculate the Curie
temperatures for the Laves lattices from the relation:
\begin{equation}
T_{c}^{*}=-\frac{2J/k_{B}}{\log v_{c}}\,,\label{9}
\end{equation}
which follows from duality, as shown in  \cite{Kramers_Wannier_1941}. The values
of the Curie temperatures for the dual lattices $\Lambda^{*}$ are
reported in collum 5 of Table 3.

\section{Conclusions}

We calculated the exact Curie temperatures for the two dimensional
Ising model on all Archimedean and Laves lattices using the combinatorial
approach of Feynman and Vodvicenko. This results are summarized in Table 3. In the particular case of the
lattices $(4,6,12)$, $(3,4,6,4)$, $(3^{4},6)$ and their duals $[4,6,12]$,
$[3,4,6,4]$, $[3^{4},6]$ our results were known in the literature
only approximately from Monte Carlo simulations. This numerical
results agree, within errors, with our analytical findings. The universal
character of the phase transition is reflected in the singularity
of the free energy, which is logarithmic ($\alpha=0$) for all lattices.
The Curie temperatures are not universal and we checked that they
depend only on the topology of the lattice, but not on the particular
embedding of it. The table of critical temperatures we compiled
can be a useful reference to future numerical and theoretical work
on the Ising model on a large class of two dimensional lattices.

\section*{Acknowledgments}

I would like to thank Prof. K. Binder for reading the manuscript and
for useful discussion and suggestions.

\end{document}